# A proposal of a Monte Carlo renormalization group transformation

L. A. Fernández [a], A. Muñoz Sudupe [a], J.J. Ruiz-Lorenzo [b] and A. Tarancón [c]

[a]Departamento de Física Teórica, Universidad Complutense de Madrid, 28040 Madrid, Spain

[b] Dipartimento di Fisica, Università di Roma I, P. A. Moro 2, 00185 Roma, Italy

[c]Departamento de Física Teórica, Universidad de Zaragoza, 50009 Zaragoza, Spain.

We propose a family of renormalization group transformations characterized by free parameters that may be tuned in order to reduce the truncation effects. As a check we test them in the three dimensional XY model. The Schwinger–Dyson equations are used to study the renormalization group flow.

## 1. INTRODUCTION

One of the key problems of the application of the renormalization group (RG) ideas in the context of Monte Carlo Simulations is the growth, under the RG transformations, of significant couplings not considered in the original action. This effect being a consequence of a RG fixed point that is situated relatively far away from the simulation point. Our proposal will be to improve one of the ingredients of the RG transformation, namely the field renormalization definition, in such a way that it will bring the RG fixed point closer to the simulation point so as to reduce the truncation effects. At the same time we want to keep the RG transformation as simple as possible because we are mainly interested in gauge theories. In order to test the reliability of the method we have chosen the three dimensional XY model, with well known critical properties.

Our procedure will be to generate configurations at the critical point $\beta_{1,c} = 0.45420(2)$, where the appropriate observables will be measured. We will then perform iteratively the RG transformations for different parameters and measure the observables on the blocked lattices. As we will see it is possible, through the lattice Schwinger–Dyson equations [1], to obtain from these measures the (blocked) renormalized couplings at each blocking level. We are thus able to follow the RG flow in the two dimensional parameter space of couplings $\beta_1, \beta_2$, nearest neighbors and next to nearest neighbors, respectively. We compute also the thermal exponent $\nu$ from the RG flow and from a Finite Size Scaling Analysis.

## 2. RG TRANSFORMATION

A RG transformation consists on renormalizing the fields and blocking the lattice from $L$ to $L/2$. For gauge theories one is forced, in order to preserve gauge invariance, to take the mean over ordered products of links along fixed–ends trajectories. This procedure can become hard to compute specially for parallel machines. Our approach would be to perform a smearing (relaxation) step previous to the actual field renormalization (decimation). These relaxational techniques are well known in other problems like spectroscopy or topological studies [2].

There are different ways to define the smearing step, a plausible guess would be to consider the relaxation driven by the heat equation

$$\frac{\partial \varphi(x,\tau)}{\partial \tau} = -\frac{\delta \mathcal{H}}{\delta \varphi(x,\tau)}.$$

However, at finite lattice spacing it is not obvious which actual transformation performs better. We will choose a smearing transformation that *a posteriori* will show to have better properties. The criterion that we impose to deduce it is that it should preserve locally the form of the Hamiltonian. For the XY model for instance, the contribution of the spin at site $n$ to the energy can be written as $\text{Re}\{e^{i\theta_n} \sum_{\pm\mu} e^{-i\theta_{n+\mu}}\} = \text{Re}\rho e^{i\alpha}$, $\rho >$

0. With the transformation

$$\theta_{n,s+1} = \theta_{n,s} + \epsilon \mathrm{Arg} \sum_{\pm\mu} e^{i(\theta_{n+\mu} - \theta_n)}$$

it is easily seen that, locally, the energy changes to $\rho \cos[(1-\epsilon)\alpha]$. Other options have been tested and we have found that they are less efficient.

The smearing parameter $\epsilon$ can be combined with the iterative number of smearing steps $n_s$ in order to select the RG transformation that has its fixed point as close as possible to the simulation (critical) point.

We use the S–D equations to follow the RG flow in the coupling space. They give us a linear systems of equations at every blocking level. For the XY model, assuming that the interaction is exclusively nearest neighbor $\beta_1 \neq 0$ at all blocking levels the S–D may be written as

$$\beta_1^b = \frac{\left\langle \sum_{\pm\mu} \cos(\theta_n^b - \theta_{n+\mu}^b) \right\rangle}{\left\langle (\sum_{\pm\mu} \sin(\theta_n^b - \theta_{n+\mu}^b))^2 \right\rangle} \ . \quad (1)$$

Should we consider also next to nearest neighbors interaction, the S–D equations would provide a $2 \times 2$ system of linear equations (see [3] for details).

We have mainly used the Wolff's Single Cluster (WSC) algorithm [4] with lattice sizes $L = 8, 16, 32$ and $64$, where we have generated 100, 50, 40 and 10 thousands of configurations respectively. Successive configurations are separated by a mean of 200 single cluster spin updates and one sweep of Metropolis. We have the interesting opportunity to study the performance of this updating algorithm at different scales and to compare it with others such as the Metropolis algorithm see Fig.1. In it we have plotted the integrated autocorrelation time for the pure WSC, Metropolis and the one used in this paper respectively. There it can be seen that, as expected, the WSC performs better at large scales while Metropolis do the opposite. Notice also that by adding just one Metropolis step to every 200 WSC updates all the scales are similarly treated.

The results presented amount for a total CPU time of only one month of Workstation. We remark that our interest is mainly on the RG trans-formation itself and that we apply it to the $3d$ XY model only as a test.

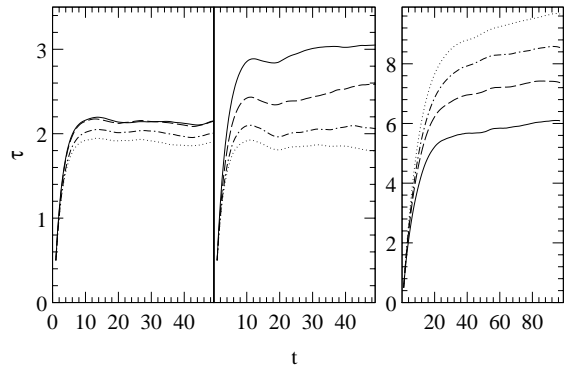

Figure 1. Comparation of the integrated autocorrelation times for from left to rigth: 1 Metropolis for each 200 Wolff's Single Cluster update, only Wolff's Single cluster and Metropolis. The solid line corresponds to $L = 16$, the dashed to $L = 8$, the dashed-dotted to $L = 4$ and the dotted to $L = 2$. We plot the integral of the normalized autocorrelation between 0 and $t$.

As a first test of the efficiency we try to tune the smearing parameters in such a way that the fixed point of the so defined RG transformation would be as close as possible to $\beta^* = (\beta_c, 0, 0, ...)$. To do that we compute the blocked couplings $\beta_1^b$, calculated with (1) in a $16^3$ lattice for different smearing parameters choices. We obtained that the matching of the blocked couplings is optimal for the RG defined by $(n_s, \epsilon) = (2, 0.285)$.

As a second test it is possible, using a two dimensional generalization of (1), to compute the nearest and next to nearest neighbors couplings $\beta_1, \beta_2$ at blocking level $b$. Thus we can draw the RG flow in the two dimensional parameter space $(\beta_1, \beta_2)$ (see Fig. 2). There it is clearly seen the importance of the appropriate tuning of the RG parameters $(n_s, \epsilon)$. For the choice $(2, 0.285)$ it can be seen that in the first two RG steps the flow approaches the fixed point $\approx (0.43, 0.02)$ while in the third step and, more obviously, in the fourth step it leaves the fixed point in the relevant direction, following the renormalized trajectory. The difference between the coupling $\beta_1^b$ computed in the one coupling (nearest neighbors) approximation and when two couplings are considered, in a $64^3$, is small (5%) for the first four levels of coupling giving us an estimation of the order of the systematic errors introduced. Strong finite size effects manifest themselves in a higher difference between the two calculations of the coupling in the last blocking level.

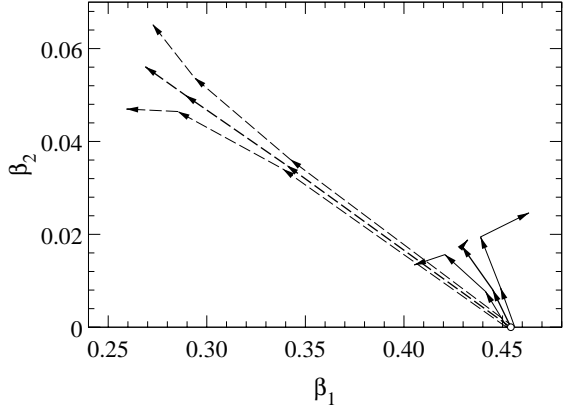

Figure 2. Flow of the Renormalization Group transformations in the two dimensional coupling space for a $32^3$ lattice. The solid lines correspond to the selection $(n_s = 2, \epsilon = 0.285)$, the dashed ones to $(n_s = 2, \epsilon = 0.2)$.

We have also computed the critical exponent $\nu$ from derivatives of the couplings, obtaining values with systematic errors lower than 3% just after one RG step. For the $L = 64$ lattice we obtain from a linear fit $\nu = 0.0649(20)$.

Another somewhat independent measure can be obtained from Finite Size Scaling arguments. Starting from different lattice sizes $L_1, L_2$ we arrive, after several RG steps, to the same given size $L$ (usually $L = 2$). By comparing the $\beta$-derivatives of some observable on the $L = 2$ lattice coming from $L_1, L_2$, we obtain directly the critical exponent through

$$\frac{1}{\nu} = \frac{\log\left(\frac{d}{d\beta}\langle O \rangle_{L_1,\beta} / \frac{d}{d\beta}\langle O \rangle_{L_2,\beta}\right)\Big|_{\beta=\beta_c}}{\log \frac{L_1}{L_2}}. \quad (2)$$

In Fig. 3 we plot the estimation using the previous formula for the plaquette operator as well as the $\beta_1$ coupling obtained from the S–D equations.

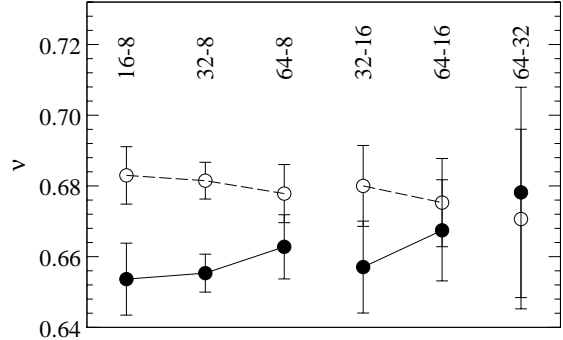

Figure 3. Critical exponent $\nu$ obtained from Finite Size Scaling analysis of the plaquette operator (filled circles) and $\beta_1$ coupling respectively.

In summary we have proposed a family of RG transformations characterized by the parameters $(n_s, \epsilon)$ whose main advantages are: it is easily implemented, has small truncation effects (at least for the tested model) and it may be ported with small modifications to parallel machines.

We acknowledge CICyT for partial financial support with the projects AEN93-0604, AEN93-0776 and AEN94-218. One of us (JJRL) is granted by MEC (Spain).